\documentstyle[osa,manuscript,epsbox]{revtex}  % DON'T CHANGE %

\begin{document}                % INITIALIZE - DONT CHANGE % %  %

\title{Liquid-Solid Phase Transition of the System with Two particles in a
Rectangular Box.}

\author{Akinori Awazu\footnote{E-mail: awa@zenon.ms.osakafu-u.ac.jp}\\ }

\address{Department of Mathematical Sciences\\
Osaka Prefecture University, Sakai 599-8531 Japan.         } %

\maketitle
\begin{abstract}
We study the statistical properties of two hard spheres in a two
dimensional rectangular box. In this system, the relation like 
Van der Waals equation loop is obtained between the width of the box and
the pressure working on side walls. The auto-correlation function of
each particle's position is calculated numerically. By this
calculation near the critical width, the time at which the
correlation become zero gets longer according to the increase of the height
of the box. Moreover, fast and slow relaxation processes
like $\alpha$ and $\beta$ relaxations observed in supper cooled liquid 
are observed when the height of the
box is sufficiently large. These relaxation processes are discussed with the
probability distribution of relative position of two particles.
\vspace{2mm}
PACS number(s):\\

\end{abstract}

Liquid-solid phase transition is one of the most familiar phenomena
for us. This transition of a system with so many degrees of
freedom has been studied through many kinds of analytical models and numerical
models\cite{re1}.
In numerical parts, the liquid-solid phase transition is observed in
the system containing $10^{1 \sim 4}$ hard or soft core particles
by Monte-Calro simulation and Molecular Dynamics
simulation\cite{re21,re2,re211,re22,re31}. The motions of individual
particles (moleculars)
are different between in liquid phase and in solid phase as follows.   
In the liquid phase, particles can exchange their own
positions each other, and each particle can wander all around the system.
On the other hand, in the solid phase, particles cannot exchange their
own positions, and they move only restricted small areas.

Now we consider a rectangular box containing two hard spheres with
same diameter $d$. The height of the box is larger than $2d$.
When the width of the box is larger than $2d$, these spheres can
exchange their own positions (Fig.1 (a)). However, these particles
cannot exchange their positions when the width of box is smaller than
$2d$ (Fig.1 (b)). From these facts, we regard the former state and the
latter state as the simplest forms of, respectively, the liquid state
and the solid state.
Then, a problem arises; In such a simple system near the critical
width ($=2d$) of the box, can we find characteristic phenomena like Alder
transition\cite{re2,re211,re31} of the system with many hard spheres ?
In this paper, we focus on statistical and dynamical properties of the 
spheres near the critical width of the box to understand this problem. 

The system on our consideration consists of two dimensional hard sphere
particles with unit mass and unit radius which are confined in a
two-dimensional rectangular box. Here, the width and the height of the box
are respectively $a$ and
$b$, and all the walls are rigid (Fig.1). The interaction between
two particles or between a particle and a wall, is only through hard core
collisions. These tasks are implemented in the following manner; the
tangential velocities to the collision plane are preserved, while the normal
component of relative velocity $\Delta v_{n}$ changes into $-\Delta v_{n}$.
The total energy of this system is given as $1$. Because this system
consists of rigid spheres and rigid walls, the qualitative
behaviors are independent of the total energy. We set $b>4$ for most
of our discussions which means these two particles can exchange
their positions in horizontal direction with each other.
Note, in this paper, the state with $a>4$ is regarded as
the liquid state and the state with $a<4$ is regarded as the solid
state.

Figure 2 (a) and (b) are typical relations between the width $a$ and the
pressure $P_{a}$ which works on side walls for each height $b=4.3 \sim
6.7$. Here, $P_{a}$ is defined as the time average of the
impulses caused by the bouncing of particles on side walls per unit length per 
unit time.
In these figures, the region of $a$, in which the volume compressibility
is negative, appears for small $b$ around the critical width $a^{*}=4.0$.
These curvatures are similar to the Van der Waals loop\cite{re1} or
the loop of Alder transition\cite{re2,re211,re31} which includes the
liquid-solid coexisting region. This negative volume compressibility
seems to indicate the appearance of phase transition around the
critical width $a^{*}=4.0$ which distinguishes the solid state from the
liquid state. If $b$ becomes larger than a critical value $b^{*}
\sim 6.0$, however, the form of this curvature is loosened, and the
compressibility becomes positive for all $a$. In this case, we cannot
observe the distinction between liquid state and solid state. 
Figure 2 (c) and (d) are the typical relations between the width $a$ and the
pressure which works on upper and lower walls $P_{b}$ for each height
$b=4.3 \sim 6.7$. Here, $P_{b}$ is defined as the time average of the
impulses caused by the bouncing of particles on upper and lower walls per
unit length per unit time. Different from the relation between $a$ and $P_{a}$,
$P_{b}$ decrease monotonically with the increase of $a$. Such
anisotropy seems one of the characteristic features of this system,
which doesn't appear in the system with many hard
spheres. If we focus only on the $a$-$P_{a}$ 
relation, however, this system can be regarded as one of the simplest
model to imitate phenomena of the liquid-solid phase transition.
Here, $a$ and $P_{a}$ correspond
to, respectively, the volume and pressure of the system with many particles.

Now, we consider the counterpart of the above $b$ in the system with many
hard spheres or in more general systems with many degrees of
freedom. Figure 3 (a)
is a typical auto-correlation functions of the position of each particle 
$C(t)=<{\bf x}(0){\bf x}(t)>/<{\bf x}(0){\bf x}(0)>$ in the solid
state ($a=3.8$) and liquid
states ($a=4.1$, $a=4.5$, and $a=5.0$ with $b=5.5$). Here, ${\bf x}(t)$
is the position of a particle.
In the liquid state, the relaxation process becomes slower as the
width comes closer to
the critical value. In the solid state ($a=3.8$), the correlation
function has a finite value for $t \to \infty$ because two particles
cannot exchange their positions. 
Figure 3 (b) is the auto-correlation functions for $b=4.3, 5.1, 5.9,
6.7$ near the critical width $a^{*}$ ($a=4.1$).
These curvatures indicate that each relaxation process contains the fast
and the slow relaxation processes for a little over the critical width.
These fast and slow relaxations can be fitted with functions,
respectively, $t^{-\beta}$ ($\beta \sim 0.75$) 
and $exp(-t^{\alpha})$ ($\alpha \sim 0.67$). By increasing $b$, the
form of $C(t)$  changes as follows. i)The time at which the correlation
becomes zero gets longer. ii)When $b$ is larger
than a critical value $b^{**}\sim 5.1$, the fast relaxation and the
slow relaxation are separated clearly by the appearance of plateau.
These relaxations are respectively similar to $\beta$ and $\alpha$
relaxations of the density fluctuations in super cooled
liquid\cite{re3}. The system which includes non-uniform moleculars
tends to become super cooled liquid state when it is cooled or
compressed\cite{re3}. Moreover, the liquid-solid
coexisting region disappears in the system with many hard
spheres when the size polydispersity of spheres is 
larger than a critical value\cite{re31}. From these facts, we expect
that present two particles system imitates the phase transition
of a system which consists of many non-uniform
elements. Here, the quantity $b$ corresponds to the dispersion of particles'
characters like the size polydispersity in the system with many particles.
In order to discuss the mechanism of above simulation results, we
focus on the statistical properties of each particle's trajectory for
each $b$ near the width $a^{*}$.

Figure 4 (a), (b) and (c) are typical trajectories of centers of
particles for, respectively, $b=4.7$, $b=5.7$ and $b=6.7$ with $a=4.1$.
If the volume of the box is enough large and we can ignore the particles'
volume, these trajectories fill the rectangular region which is
enclose by points $(\frac{a-d}{2},\frac{b-d}{2})$,
$(\frac{a-d}{2},-\frac{b-d}{2})$, $(-\frac{a-d}{2},\frac{b-d}{2})$ and
$(-\frac{a-d}{2},-\frac{b-d}{2})$
(Fig.4(c)). Here, $d$ is the diameter of each particle which is set
as 2 in our discussion. This means particles wander all around the box
like ideal gas systems. 
When the size of box becomes small, however, the finite volume effect 
of particles become evident.
Particularly, if the relation  
\begin{equation}
(a-d)^{2}+(b-d)^{2}<(2d)^{2}
\end{equation}
is satisfied, inhibited region for particles' center to come in
appears around the central part of the box.
When $b<2+2\sqrt{2}$ is satisfied, the above
equation is satisfied around $a=a^{*}=4$ and the trajectory of the
center of particle is given like Fig.4 (a). This is also the case for
$a<a^{*}=4$ where the trajectory of the center of particle
is given like Fig.4 (b) if $2+2\sqrt{2}<b< b^{*}=6$.
In these cases, each trajectory of the center of particle is similar to
what is observed in Sinai billiard\cite{re4}.
Thus, in order to discuss these situations, we consider 
the $(a-d) \times (b-d)$ rectangular Sinai billiard which includes a
hard sphere with diameter $d$ ($d=2$). Using the equi-partition
rule, we can calculate the $a$-$P_{a}$ relation of Sinai billiard 
analytically like correlated cell
model\cite{re5}. For each particle in Sinai billiard, the entropy $S$
is given by using the phase space volume $S=log(A_{r}(a,b)A_{v})$ and free
energy $F$ is given as $F=U-S$. (Boltzmann constant $k_{B}$
and temperature $T$ are set as 1.) Here, $A_{r}(a,b)$ and $A_{v}$ are
phase space volumes of, respectively, real space part and velocity space
part, and internal energy $U$ is constant because this system consists 
of hard walls and a hard sphere. $A_{r}(a,b)$ is given as
\begin{equation}
A_{r}=(a-d)(b-d)-\pi \cdots (a > 4)
\end{equation}
\begin{equation}
A_{r}=(a-d)(b-d)-(a-d)cos\theta-2\theta \cdots
(a \le 4)
\end{equation}
where $sin\theta=\frac{a-d}{d}$ and $d=2$.  Using the above relations
also with $P_{a}b=-\frac{\partial F}{\partial a}$, $a$-$P_{a}$ relation of the 
system for $b<6$ is obtained, and we can observe the liquid-solid
phase transition like Fig. 5(a).
In this case, the width $a=a^{*}=4$ is a singular point and this point 
gives the maximal pressure independent of $b$. 
On the other hand, however, the $a$-$P_{a}$ relation in simulation
results (Fig.2 (a) and (b)) have the inflection point near the critical width
$a=a_{*}=4$, and the form of each curvatures is smooth. 
In addition, we consider a square box system in which $a$ and $b$ are
varied with $a=b$. This system satisfies $b<2+2\sqrt{2}$ around $a
\sim a^{*}$. 
In the same way of the calculation of above rectangular Sinai billiard, the
$a$-$P_{a}$ relation of the $(a-d) \times (a-d)$ square Sinai
billiard can be obtained analytically. In this calculation, the
profile of the $a$-$P_{a}$ relation of the square Sinai
billiard is obtained
as the almost same one as the results of the rectangular billiard.
However, it is remarkable that $P_{a}$($=P_{b}$) decreases monotonically with
increasing $a$($=b$) in the simulation of two particles in the square
box (Fig.5 (b)).

Finally, we discuss the mechanism of the appearance of plateau in
$C(t)$, the auto-correlation function of each particle's position, near
the width $a=a^{*}$ considering  statistical properties of particles' trajectories. 
Now, we define $Pro_{\delta x}$ and
$Pro_{\delta y}$ as the probability distributions of $\delta x$ and
$\delta y$. Here, $\delta x$ is horizontal component and $\delta y$ is 
vertical component of relative position vector between centers of two
particles. 
Figure 6 (a) and (b) are respectively $\delta y$-$Pro_{\delta y}$
relations and $\delta x$-$Pro_{\delta x}$ relations of the case $b=4.3,
4.7, 5.1, 5.5, 5.9$ with $a=4.1$.
In Fig.6 (a), the maximum points of $Pro_{\delta y}$ are always far from the
point $\delta y = 0$. In Fig.6 (b), however,
the position of the maximum point of $Pro_{\delta x}$ depends on $b$. 
When $b$ is small, the maximum points of $Pro_{\delta x}$ are far from 
the point $\delta x = 0$. This means that two particles tend to face each
other on a diagonal line of box.
By this tendency, it is rather easy for these two particles to
exchange their positions both in vertical and horizontal direction.
Hence, the time at which $C(t)$ become $0$
is relatively short. On the other hand, only one maximum point of 
$Pro_{\delta x}$ appear at $\delta x=0$ when $b$
is larger than the critical value $b^{**} \sim 5.1$.
In this case, the height of box is enough large to
change particle's positions almost freely but only in the horizontal direction.
This is considered as the origin of fast relaxation in the simulation.
This situation also means that two particles, in statistical sense,
tend to line up on a vertical line.
In this case, the exchange of two particles' positions in
vertical direction is cut across strongly. This fact originates the
appearance of the plateau in $C(t)$, after which the slow relaxation
starts. 

In this paper, liquid-solid phase transition and the long time
correlation of two hard spheres confined in a two dimensional rectangular box
is studied. Between the width of the box and
the pressure at the side walls, the relation like Van der Waals
equation is obtained. However, the range of the box width, in which the volume 
compressibility is negative, disappears when the height of this box passes 
through a critical value.
The auto-correlation function of each particle's position is
calculated near the critical width. According to the increase of the height
of this box, the time at which this correlation become zero gets
longer. Moreover, a fast relaxation and a
slow relaxation are separated clearly by the appearance of plateau when
the height of this box is sufficiently large. These relaxation
processes are discussed considering the form of the probability
distribution of relative position of two particles.
As a conclusion, this system is considered to be one of the simplest
system which imitates the liquid-solid phase transition of the system
which has been believed to be a characteristic nature of systems with
many non-uniformed elements.
Still, for the relation between the
width of the box and the pressure at the side walls, some 
discrepancies appear between the analytical and simulation results.
Thus further consideration is required on the 
dynamical properties like the long time correlation which forbids the
equi-partition. These topics seem to have strong relation with the
slow dynamics in
Hamilton dynamical systems\cite{re6}. In addition, the pressure which works
on walls is anisotropic in our system, while the pressure of the
system with many particles is usually uniform. This also is a problems
to be solved. Moreover, the understanding of the glass transition or the other
non-equilibrium system\cite{re8} through our simple model is one of future issues.

The author is grateful to H.Nishimori, N.Ito, S.Sasa, H.Hayakawa,
M.Sano and K.Sekimoto for useful discussions. This research was
supported in part by the Ibaraki University SVBL and Grant-in-Aid for
JSPS Felows 10376.

\newpage

\begin{figure}[h]
\caption[]{Illustration of two particles system in rectangular
box. (a)Width of box is larger than sum of two diameters (Liquid
state), and (b)width of box is smaller than sum of two diameters
(solid state).}
\end{figure}

\begin{figure}[h]
\caption[]{Relations between the width $a$ and the
pressure $P_{a}$ with (a)$b=4.3, 4.5, 4.7, 4.9$ (b) $b=5.1, 5.5, 5.9, 6.3,
6.7$ in descending order, and that between the width
$a$ and the pressure $P_{b}$ with (c)$b=4.3, 4.5, 4.7, 4.9$ (d)
$b=5.1, 5.5, 5.9, 6.3, 6.7$.}
\end{figure}

\begin{figure}[h]	
\caption[]{Auto-correlation function of each particle's position
$C(t)$, respectively, (a)$a=3.8, 4.1, 4.5, 5.0$ with $b=5.5$, and
(b)$b=4.3, 5.1, 5.9, 6.7$ with $a=4.1$. Fitting lines are $\propto
t^{-0.75}$ and $\propto exp(-t^{0.67})$.}
\end{figure}

\begin{figure}[h]
\caption[]{Typical trajectories of particles
respectively (a)$b=4.7$, (b)$b=5.7$ and (c)$b=6.7$ with $a=4.1$.}
\end{figure}

\begin{figure}[h]
\caption[]{(a)$a$-$P_{a}$ relations of rectangular Sinai billiard with
$b=4.5, 5.1$ and $5.7$, and (b)$a$-$P_{a}$ relations of two particles
in square box.}
\end{figure}

\begin{figure}[h]
\caption[]{Probability distributions (a)$Pro_{\delta y}$ of vertical
component of relative position vector $\delta y$, and (b)$Pro_{\delta
x}$ of horizontal component of relative position vector $\delta x$
between two particles with $a=4.1$. $b=4.3, 4.7, 5.1, 5.5$
and $5.9$ from the bottom up near $\delta x=0.0$ in (b).}
\end{figure}

\end{document}